\documentclass[12pt]{amsart}

\usepackage{graphicx} 

\usepackage{amsmath}
\usepackage{amsthm}
\usepackage{amsfonts}
\usepackage{amssymb}
\usepackage{latexsym}
\usepackage{color}
\usepackage{mathtools}
\usepackage{multirow}
\usepackage{array}
\usepackage{tabularx}
\usepackage[normalem]{ulem}

\usepackage{hyperref}
\usepackage[top=1in, bottom=1.1in, left=1in, right=1in]{geometry}

\newcommand{\Z}{\mathbb{Z}}

\newcommand{\Mod}[1]{\ (\mathrm{mod}\ #1)}

\author[]{Alexander Demin}
\address{Laboratoire d'informatique de l'{\'E}cole polytechnique, LIX, UMR 7161, CNRS, 1 rue Honor{\'e} d'Estienne d'Orves, 91120 Palaiseau, France}
\email{alexander.demin@lix.polytechnique.fr}

\author[]{Alexey Ovchinnikov}
\address{Department of Mathematics, Queens College, City University of New York, Queens, NY 11367}\email{aovchinnikov@qc.cuny.edu}\thanks{Research of the second and third authors was partially supported by a CUNY Institutional Research Grant Award}

\author[]{Vladimir Shpilrain}
\address{Department of Mathematics, The City College of New York, New York,
NY 10031} \email{shpilrain@yahoo.com}

\begin{document}

\title{Redactable blockchains and polynomial equations}

\begin{abstract}
We develop new tools for constructing redactable authenticated data structures with post-quantum security. In our construction, inverting the proposed one-way function means solving a polynomial equation (or a system of polynomial equations) in more than one variable. This is presently considered quantum-safe, i.e., there is no known quantum algorithm that could solve this problem efficiently if parameters are chosen wisely.
\end{abstract}

\maketitle

\section{Introduction}\label{Introduction}

A crucial part of most artificial intelligence systems is an adaptive network of electronic devices. Perhaps the most popular and quickly proliferating class of such networks is the  {\it Internet of Things}.

The Internet of Things (IoT) represents one of the most significant technologies of
this century. It is a natural evolution of the Internet (of computers) to embedded and cyberphysical
systems, ``things" that, while not obviously computers themselves, nevertheless have
computers inside them.

IoT is experiencing exponential growth in research and industry, but
it still suffers from privacy and security vulnerabilities. Conventional security and privacy
approaches tend to be inapplicable for IoT, mainly due to its decentralized topology and the
resource-constraints of the majority of its devices. Blockchains have been recently used to provide security and privacy in peer-to-peer networks and artificial intelligence  systems such as, for example, smart cars, smart homes, smart cities, etc.

A blockchain can be either public or private.  A typical example of a public blockchain is bitcoin. In contrast, a private blockchain network is where the participants are known and trusted: for example, an industry group, or a military unit, or in fact any private network, big or small.

Originally, blockchains were created to support the bitcoin network, which is public. Immutability for such a network is crucial. More recently, with the ideas of artificial intelligence gaining momentum, private networks have taken the center stage, and this creates new challenges. In particular, it is desirable, while preserving the tampering detection property, to allow  someone with higher privileged access like a system  administrator or another authority to be able to change the data or erase (“forget") it. A blockchain that can be changed like this is called {\it redactable}. The need for a blockchain (even a public one!) to be redactable is well explained by the following quote from  The New York Times \cite{NYT}: ``That permanence has been vital in building trust in the decentralized currencies, which are used by millions of people. But it could severely limit blockchain’s usefulness in other areas of financial services relied on by billions of people. By clashing with new privacy laws like the ``right to be forgotten" and by making it nearly impossible to resolve human error and mischief efficiently, the blockchain’s immutability could end up being its own worst enemy." Eliminating redundancies and deleting outdated records is especially vital for networks (such as IoT, for instance) of computationally limited devices.

The present paper proposes a new secure, private, and lightweight architecture for redactable  blockchains based on multivariable polynomial equations.

Earlier constructions (e.g. \cite{Giuseppe, Thyagarajan, GS}) of redactable blockchains were either not practical, or not quantum-safe, or both. While preserving the simplicity and efficiency of the construction in \cite{GS}, we develop new tools for constructing redactable authenticated data structures {\it with post-quantum security}. One possible way to avoid quantum attacks based on solving the hidden subgroup problem (including the attacks in \cite{Shor}) is {\it not} to use one-way functions that utilize, in an essential way, one or another (semi)group structure. In our construction, inverting the proposed one-way function means solving a polynomial equation (or a system of polynomial equations) in more than one variable. This is presently considered quantum-safe, i.e., there is no known quantum algorithm that could solve this problem efficiently if parameters are chosen wisely.

\section{Background}

Our approach in this proposal is focused on private blockchains. It is quite different from \cite{Giuseppe,Thyagarajan, Puddu} and other methods and is transparent and easily implementable.

On a high level, our method bears
some similarity with the construction in \cite{GS} that was based on the computational hardness of the RSA problem and therefore was not quantum-safe, whereas in this paper we aim at quantum-safe constructions. Still, here we first briefly describe the construction from \cite{GS} to give a background and put things in perspective.

There are several possible structures of a redactable blockchain. Our general method should work with any known structure,
but to make  the exposition as clear as possible we choose a
simple structure as follows. Each block $B_i$ will have 3 parts: a permanent prefix $P_i$, the actual content $C_i$,  and a redactable
suffix $X_i$.  There is also a function (e.g. a hash function) $h_i=H(P_i, C_i)$, where $H$ is a public (hash) function, and a public one-way function $F$ such that $F(h_i, X_i)=P_{i+1}$. To make such a blockchain redactable, a central authority should have a private key that would allow for replacing $C_i$ with an arbitrary $C'_i$ of his/her choice, so that upon a suitable selection of the new suffix $X'_i$, the equality $F(h'_i, X'_i)=P_{i+1}$ would still hold.
\subsection*{An RSA-based construction.} Now we briefly sketch the construction from \cite{GS}.
\subsubsection*{Public information:}
\begin{itemize}
\item a large integer $n$, which is a product of two large primes
\item a hash function $H$, e.g. SHA-512. (We emphasize again that our method can be used with any reasonable hash function, so the reader can replace SHA-512 here with his/her favorite hash function.)
\end{itemize}
\subsubsection*{Private information:}
\begin{itemize}\item prime factors of $n = pq$. These $p$ and $q$ should be safe primes, as in modern implementations of RSA. A safe prime is of the form $2r + 1$, where $r$ is another prime~\cite{zur}.
\end{itemize}
\subsubsection*{Block structure.} In each block $B_i$, there is a prefix $P_i$,
the actual content $C_i$ (e.g.  a transaction description), and  
a suffix $X_i$, which is a nonzero  integer modulo $n$.
We also want $X_i$ {\it not} to have order 2, i.e., $X_i^2 \ne 1 \Mod{n}$. This will ensure that $X_i$ does not have a small order since $p$ and $q$ are safe primes, so if the order of an element of the multiplicative group $\Z^*_{pq}$ is not 2, then it is large.  Thus, whoever builds a block $B_i$, selects $X_i$ at random on integers between 1  and $n-1$  and then checks if $X_i^2 \ne 1 \Mod{n}$. If $X_i^2 = 1 \Mod{n}$, random selection of $X_i$ is repeated.
Once a proper $X_i$ is selected, a public hash function $H$ 
(e.~g.~SHA-512)
is applied to concatenation of $P_i$ and $C_i$ to produce $h_i=H(P_i, C_i)$, and  $h_i$ is then converted to an integer $d_i$ modulo $n$.   The prefix $P_{i+1}$ of the next block is then computed as $P_{i+1} = (X_i)^{d_i} \Mod{n}$.
\subsubsection*{Private redactability.} \label{redactability} Now suppose the central authority, Alice, who is in possession of the private key, wants to change the content of a block $B_i$ from $C_i$ to $C'_i$ but does not want to change any other block. Then Alice computes the hash $h'_i = H(P_i, C'_i)$ and converts it to an integer $d'_i$ modulo $n$. The number  $d_i'$ should be relatively prime to $\phi(n)$, the Euler function of $n$. If it is not, then Alice should use a padding to have $d_i'$ relatively prime to $\phi(n)$. Once it is, Alice finds the inverse $e'_i$ of $d_i'$ modulo $\phi(n)$. Then she computes $X'_i = P^{e'_i}_{i+1} \Mod{n}$. The integrity check now gives: \[(X'_i)^{d_i'} = \left(P^{e'_i}_{i+1}\right)^{d_i'} = P_{i+1} \Mod{n}\] because $e'_i d_i' = 1 \Mod{\phi(n)}$ and
$(P_{i+1})^{\phi(n)} = 1 \Mod{n}$.
\subsubsection*{Public immutability} \label{immutability} As can be seen from the ``Block structure" paragraph, a party who would like to change the content of a single block, 
must
essentially solve the RSA problem: recover $X$ from $n$,
$X^{d} \Mod{n}$, and $d$, where $d$ is relatively prime to $\phi(n)$. This is considered computationally infeasible for an appropriate choice of $n$ and a random $d, ~0 < d < n$.

\section{Basic construction}\label{plan}

Now we describe an integrity condition for blockchains that would involve quantum-safe one-way functions with trapdoor.
In this section, to illustrate the idea, we first describe a basic version of the construction. The basic version is susceptible to several attacks. Later in the text (Section \ref{sec:advanced}), we propose a construction that is resilient to those attacks.

Recall that $P_i$ denotes the prefix of the block $B_i$. Prefixes are immutable, i.e., they cannot be changed by anybody. We do not go into details here on how this immutability can be arranged; literature on blockchains is awash with that.
Then, $C_i$ denotes the content of the block $B_i$; this is the crucial part of a block where meaningful information is stored. Finally, $S_i$ denotes the suffix of the block $B_i$; this is an auxiliary part that makes it possible for a central authority to change the content $C_i$ without violating the integrity condition.

Let $t$ be a new variable. We are assuming that all $C_i$, $S_i$, and
$P_i$ are univariate polynomials in $t$ of degree $d$ with coefficients in $\mathbb{Z}_n$ for a fixed large prime $n$. We assume that $d<n$. Needless to say, any content $C_i$ can be encoded this way if $n$ is large enough, say several hundred bits.

A crucial part of any blockchain (redactable or not) is the {\it integrity condition}. In our case, the integrity condition is of the following form:
$$F(P_i, C_i, S_i) = P_{i+1} \Mod{(t-t_0) \cdots (t-t_d)}$$
where $F$ is a public polynomial and $t_0, \ldots, t_d$ are public distinct integers modulo $n$.

A trapdoor (i.e., a secret known only to a central authority) is a presentation of the polynomial $F$ in the form
$$F(x,y,z) = G(x, H(x,y) - z),$$
where $G(x, y)$ and $H(x, y)$ are private two-variable polynomials.

To change the content $C_i$ to $C_i'$ while preserving the integrity condition, it is sufficient to find a new suffix $S_i'$ such that
$H(P_i,C_i)-S_i = H(P_i,C_i')-S_i' \Mod{(t-t_0) \cdots (t-t_d)}$.
Solving this for $S_i'$ gives
$$S_i' = H(P_i,C_i')-H(P_i,C_i)+S_i \Mod{(t-t_0) \cdots (t-t_d)}.$$

Thus, whoever knows the function $H$ can change $C_i$ to any other content $C_i'$ without violating the integrity condition.
Note that any $\hat{G}, \hat{H}$ that satisfy the integrity condition can be used for this purpose (with $\hat{G}, \hat{H}$ not necessarily the same as the secret $G, H$, respectively).


\section{Possible attacks on the basic construction}\label{sec:attacks}

Now let us look at what an unauthorized party can try to do to modify $C_i$ to some $C_i'$. There are several possibilities:

\begin{enumerate}
\item[\bf Attack 1.] Solve the polynomial equation
$F(P_i, C_i', X) = P_{i+1} \Mod{(t-t_0) \cdots (t-t_d)}$
for $X$, where $X$ is a polynomial in $t$ of degree $d$ with unknown coefficients from $\Z_n$. Then this $X$ can be a new suffix of the block $B_i$. This gives rise to a square polynomial system in the coefficients of $X$. Solving such a system is a hard problem in general, but perhaps for some special polynomials it may turn out to be easy.
\medskip

\item[\bf Attack 2.] \label{item:attack-2} Recover the secret polynomial $H(x,y)$. The intruder can attempt to present $H(x,y)$ as a sum of monomials (with the degree bounded by the degree of the polynomial $F$) with unknown coefficients. Then the equation
$F(P_i, C_i', S_i') = P_{i+1} \Mod{(t-t_0) \cdots (t-t_d)}$
would translate into a system of polynomial equations in these unknown coefficients. These equations are not linear, so solving them
(even over $\Z_p$ for a prime $p$) requires Gr\"obner basis techniques as in \cite{F4,Groebner}. 
\medskip

\item[\bf Attack 3.] \label{item:attack-3} Find a functional decomposition of $F(x,y,z)$. 
The algorithm by Kozen and Landau~\cite{kozen} is efficient at finding compositions of univariate polynomials over commutative rings. The algorithm can be used in our setting by considering the multivariate polynomial $F(x,y,z)$ as a univariate polynomial in one of the variables $x,y,z$. An alternative approach~\cite{MultivariateFunctionDecomposition}, which has polynomial complexity in case the degrees  are fixed, is to find a multivariate function decomposition of $F(x,y,z)$. In our case, one could construct the secret $G(x,y), H(x,y)$ of arbitrarily large degrees, hence this algorithm would likely be inefficient.

\medskip

\item[\bf Attack 4.] \label{item:attack-4} If an attacker can observe the same block before and after it was changed, then (s)he can be in possession of the following equation:
$S_i - S_i' = H(P_i, C_i) - H(P_i, C_i') \Mod{(t-t_0) \cdots (t-t_d)}$. Here $S_i, S_i', P_i, C_i, C_i'$ are all known to the attacker.

Then, since the bound on the degree of $H(x,y)$ is known to the public, the above  equation translates into a system of linear equations in the coefficients of $H(x,y)$. Upon accumulating sufficiently many equations as above (after observing sufficiently many replacements of $C_i$ and $S_i$ by different $C_i'$ and $S_i'$, respectively), the attacker can find all the coefficients of $H(x,y)$.
\end{enumerate}

\medskip

As the experiments in Section~\ref{Experiments} show, the basic construction from Section~\ref{plan} is susceptible to each of these attacks, although the following modification of the basic construction foils Attack 4.

Specifically, the function $H(x,y)$ will appear nonlinearly in the trapdoor presentation of $F(x, y, z)$, as follows:

$$F(x,y,z) = G(x, R(H(x,y), z)),$$

\noindent where $R(H, z)$ is a private function nonlinear in the first argument and linear
in the second one. We also require that $R(H,z)$ contains a term depending only on $H$ and a term involving both $H$ and $z$.

As a simple example, take $R(H,z) = H + H^2 z +
z$, which yields
\[
F(x,y,z) = G(x, H(x,y) + (H(x,y))^2 z 
+
z).
\]

Then to change the content $C_i$ of a block $\#i$ to $C_i'$ while preserving the integrity condition $F(P_i, C_i, S_i) = P_{i+1} \Mod{(t-t_0) \cdots (t-t_d)}$, it is sufficient for a legitimate authority to find a new suffix $S_i'$ such that
\[
H(P_i,C_i) - H(P_i,C_i')  + (H(P_i,C_i)^2 + 1) S_i =  S_i' (H(P_i,C_i')^2 + 1) \Mod{(t-t_0) \cdots (t-t_d)}
\]
Since the legitimate authority knows in this equation everything except $S_i'$, this equation translates for them into a system of linear equations in the coefficients of $S_i'$, which has a unique solution whenever $H(P_i,C_i')^2 + 1$ and $(t-t_0) \cdots (t-t_d)$ are coprime, and is efficiently solvable. (Recall that $S_i'$ is a polynomial of bounded degree.) For this choice of $R(H,z)$, coprimality can be ensured by choosing $n$ so that $n = 3 \Mod{4}$, since then $-1$ is not a square modulo $n$.

At the same time, Attack 4 would yield a system of equations in the coefficients of $H(x, y)$, but this system is no longer linear.

\section{Advanced quantum-safe construction}
\label{sec:advanced}

We now propose a construction (an integrity condition and a trapdoor) that is not vulnerable to previously mentioned attacks.

Let $F(x,y,z)$ be a three-variable polynomial with coefficients given by integers modulo $n$. Introduce new indeterminates $\bar{A} = (A_0,\ldots,A_d)$, and similarly for $\bar{B}, \bar{C}, \bar{D}$. Let $\hat{P_i}(t), \hat{C_i}(t), \hat{S_i}(t), \hat{P}_{i+1}(t)$ be univariate polynomials of degree $d$ defined as follows:
\[
\begin{aligned}
    \hat{P_i}(t) = A_0 + \ldots + t^{d}A_d, \quad& \hat{C_i}(t) = B_0 + \ldots + t^{d}B_d,\\
    \hat{S_i}(t) = C_0 + \ldots + t^{d}C_d, \quad& \hat{P}_{i+1}(t) = D_0 + \ldots + t^{d}D_d.\\
\end{aligned}
\]

Note that if one specializes $\bar{A}, \bar{B}, \bar{C}, \bar{D}$ to some particular numbers, then $\hat{P_i}(t), \hat{C_i}(t), \hat{S_i}(t)$, and $\hat{P}_{i+1}(t)$ become concrete prefix, content, and suffix, that is, $P_i, C_i, S_i, P_{i+1}$.
Therefore, an integrity condition on $P_i, C_i, S_i, P_{i+1}$ can be formulated as a condition (equation) on $\bar{A}, \bar{B}, \bar{C}, \bar{D}$.




The advanced integrity condition is a system of (fully expanded) polynomial equations in $\bar{A}, \bar{B}, \bar{C}, \bar{D}$:
\[\begin{aligned}
    F(\hat{P_i}(t_0), \hat{C_i}(t_0), \hat{S_i}(t_0)) &= \hat{P}_{i+1}(t_0),\\
    &\ldots\\
    F(\hat{P_i}(t_d), \hat{C_i}(t_d), \hat{S_i}(t_d)) &= \hat{P}_{i+1}(t_d),
\end{aligned}\]
where $t_0, \ldots, t_{d}$ are private distinct integers modulo $n$. To verify the condition for some particular choice of $P_i, C_i, S_i, P_{i+1}$, one checks if the system obtained by substituting the coefficients in $\bar{A}, \bar{B}, \bar{C}, \bar{D}$ is consistent. 

Let us show how the public can find $P_{i+1}$ from  $P_i, C_i, S_i$. Consider the system of equations that defines the integrity condition; the system depends on $\bar{A}, \bar{B}, \bar{C}, \bar{D}$. Recall that $\bar{D}$ are the coefficients of $P_{i+1}$ and note that the system is linear in $\bar{D}$ by construction. Therefore, once $\bar{A}, \bar{B}, \bar{C}$ are specialized to particular values, the values of $\bar{D}$ can be obtained by solving a square linear system with $d+1$ equations.

As in the basic construction, a trapdoor is a presentation of $F$ in the form
$
F(x,y,z) = 
G(x, R(H(x,y), z))
$
The procedure for changing the content $C_i$ to $C_i'$ while preserving the integrity condition also remains the same. The only difference is that the integers $t_0,\ldots,t_d$ are private and known only to the central authority.

Then, compared to the basic construction in Section~\ref{plan}, in the construction in this section the polynomial $F(x,y,z)$ is private.
This makes direct application of Attack~2 and Attack~3 impossible. 
Additionally, as we shall see, since $F(x,y,z)$ is private, it is not possible to use shortcuts in the computation of Gr{\"o}bner bases in Attack~1.  

%
%


\medskip

\paragraph{\bf Example} For simplicity, in this example we assume that $F(x,y,z) = G(x, H(x,y)-z)$.
Calculations in this example are performed modulo $n= 11587$. Let the private polynomials $G(x,y),H(x,y)$ be
\[
\begin{aligned}
&G(x, y) = x y + x + y + 3,\\
&H(x, y) = x y + 5.
\end{aligned}
\]
Then the private $F(x,y,z)$ is
\[
F(x,y,z) = x^{2} y + x y - x z + 6 x - z + 8.
\]
Let $d = 2$. Then $\bar{A} = (A_0, A_1, A_2), \bar{B} = (B_0, B_1, B_2), \bar{C} = (C_0, C_1, C_2), \bar{D} = (D_0, D_1, D_2)$. We therefore have
\[
\begin{array}{cc}
\hat{P_i}(t) = A_0 + A_1 t + A_2 t^2 & \hat{C_i}(t) = B_0 + B_1 t + B_2 t^2\\
\hat{S_i}(t) = C_0 + C_1 t + C_2 t^2& \hat{P}_{i+1}(t) = D_0 + D_1 t + D_2 t^2
\end{array}
\]

Let $t_0,t_1,t_2 = 1,2,3$, respectively. To obtain an integrity condition for the advanced construction, we consider the following system of equations:
\[\begin{aligned}
    F(\hat{P_i}(t_0), \hat{C_i}(t_0), \hat{S_i}(t_0)) &= \hat{P}_{i+1}(t_0),\\
    F(\hat{P_i}(t_1), \hat{C_i}(t_1), \hat{S_i}(t_1)) &= \hat{P}_{i+1}(t_1),\\
    F(\hat{P_i}(t_2), \hat{C_i}(t_2), \hat{S_i}(t_2)) &= \hat{P}_{i+1}(t_2).
\end{aligned}\]
By expanding the terms, we obtain equations that depend only on $\bar{A},\bar{B}, \bar{C}, \bar{D}$; these equations give us the public integrity condition:
\small
\[\begin{aligned}
0 &= A_{0}^{2} B_{0} + A_{0}^{2} B_{1} + A_{0}^{2} B_{2} + 2 A_{0} A_{1} B_{0} + 2 A_{0} A_{1} B_{1} + 2 A_{0} A_{1} B_{2} + 2 A_{0} A_{2} B_{0} + 2 A_{0} A_{2} B_{1}\\
&+ 2 A_{0} A_{2} B_{2} + A_{0} B_{0} +A_{0} B_{1} + A_{0} B_{2} - A_{0} C_{0} - A_{0} C_{1} - A_{0} C_{2} + 6 A_{0} + A_{1}^{2} B_{0} + A_{1}^{2} B_{1}\\ 
&+ A_{1}^{2} B_{2} + 2 A_{1} A_{2} B_{0} + 2 A_{1} A_{2} B_{1} + 2 A_{1} A_{2} B_{2} + A_{1} B_{0} + A_{1} B_{1} + A_{1} B_{2} - A_{1} C_{0} - A_{1} C_{1}\\
&- A_{1} C_{2} + 6 A_{1} + A_{2}^{2} B_{0} + A_{2}^{2} B_{1} + A_{2}^{2} B_{2} + A_{2} B_{0} + A_{2} B_{1} + A_{2} B_{2} - A_{2} C_{0} - A_{2} C_{1}\\ 
&- A_{2} C_{2} + 6 A_{2} - C_{0} - C_{1} - C_{2} - D_{0} - D_{1} - D_{2} + 8\\
0 &= A_{0}^{2} B_{0} + 2 A_{0}^{2} B_{1} + 4 A_{0}^{2} B_{2} + 4 A_{0} A_{1} B_{0} + 8 A_{0} A_{1} B_{1} + 16 A_{0} A_{1} B_{2} + 8 A_{0} A_{2} B_{0} + 16 A_{0} A_{2} B_{1}\\
&+ 32 A_{0} A_{2} B_{2} + A_{0} B_{0} + 2 A_{0} B_{1} + 4 A_{0} B_{2} - A_{0} C_{0} - 2 A_{0} C_{1} - 4 A_{0} C_{2} + 6 A_{0} + 4 A_{1}^{2} B_{0} + 8 A_{1}^{2} B_{1}\\ 
& + 16 A_{1}^{2} B_{2} + 16 A_{1} A_{2} B_{0} + 32 A_{1} A_{2} B_{1} + 64 A_{1} A_{2} B_{2} + 2 A_{1} B_{0} + 4 A_{1} B_{1} + 8 A_{1} B_{2} - 2 A_{1} C_{0}\\
&- 4 A_{1} C_{1} - 8 A_{1} C_{2} + 12 A_{1} + 16 A_{2}^{2} B_{0} + 32 A_{2}^{2} B_{1} + 64 A_{2}^{2} B_{2} + 4 A_{2} B_{0} + 8 A_{2} B_{1} + 16 A_{2} B_{2}\\
&- 4 A_{2} C_{0} - 8 A_{2} C_{1} - 16 A_{2} C_{2} + 24 A_{2} - C_{0} - 2 C_{1} - 4 C_{2} - D_{0} - 2 D_{1} - 4 D_{2} + 8\\
0 &= A_{0}^{2} B_{0} + 3 A_{0}^{2} B_{1} + 9 A_{0}^{2} B_{2} + 6 A_{0} A_{1} B_{0} + 18 A_{0} A_{1} B_{1} + 54 A_{0} A_{1} B_{2} + 18 A_{0} A_{2} B_{0} + 54 A_{0} A_{2} B_{1}\\ 
&+ 162 A_{0} A_{2} B_{2} + A_{0} B_{0} + 3 A_{0} B_{1} + 9 A_{0} B_{2} - A_{0} C_{0} - 3 A_{0} C_{1} - 9 A_{0} C_{2} + 6 A_{0} + 9 A_{1}^{2} B_{0} + 27 A_{1}^{2} B_{1}\\
&+ 81 A_{1}^{2} B_{2} + 54 A_{1} A_{2} B_{0} + 162 A_{1} A_{2} B_{1} + 486 A_{1} A_{2} B_{2} + 3 A_{1} B_{0} + 9 A_{1} B_{1} + 27 A_{1} B_{2} - 3 A_{1} C_{0}\\
&- 9 A_{1} C_{1} - 27 A_{1} C_{2} + 18 A_{1} + 81 A_{2}^{2} B_{0} + 243 A_{2}^{2} B_{1} + 729 A_{2}^{2} B_{2} + 9 A_{2} B_{0} + 27 A_{2} B_{1} + 81 A_{2} B_{2}\\ 
&- 9 A_{2} C_{0} - 27 A_{2} C_{1} - 81 A_{2} C_{2} + 54 A_{2} - C_{0} - 3 C_{1} - 9 C_{2} - D_{0} - 3 D_{1} - 9 D_{2} + 8
\end{aligned}\]

Now suppose the public would like to check if the integrity condition holds for the following block:
\[
\begin{array}{c}
P_i(t) = 6533 + 9560 t + 2512 t^2\\
C_i(t) = 8147 + 6463 t + 2498 t^2\\
S_i(t) = 3507 + 5721 t + 3086 t^2\\
P_{i+1}(t) = 7891 + 7055 t + 9071 t^2
\end{array}
\]
To perform the check, one makes the assignment
\[
\begin{array}{c}
(A_0, A_1, A_2) \coloneqq (6533, 9560, 2512)\\
(B_0, B_1, B_2) \coloneqq (8147, 6463, 2498)\\
(C_0, C_1, C_2) \coloneqq (3507, 5721, 3086)\\
(D_0, D_1, D_2) \coloneqq (7891, 7055, 9071)\end{array}
\]
and checks if the system of equations in the integrity condition is consistent. In this example, all three equations become $0 = 0$ modulo $n$, and one concludes that the integrity condition is satisfied for this block.

Using this example, let us show how the public can construct $P_{i+1}(t)$ from the knowledge of $P_i(t), C_i(t), S_i(t)$. To do this, one makes the assignment
\[
\begin{array}{c}
(A_0, A_1, A_2) \coloneqq (6533, 9560, 2512)\\
(B_0, B_1, B_2) \coloneqq (8147, 6463, 2498)\\
(C_0, C_1, C_2) \coloneqq (3507, 5721, 3086)\\
\end{array}
\]
and the system of equations in the integrity condition becomes a linear system in $D_2, D_1, D_0$:
\[
\begin{aligned}
0 &= -D_0 - D_1 - D_2 + 843\\
0 &= -D_0 - 2 D_1 - 4 D_2 + 350\\
0 &= -D_0 - 3 D_1 - 9 D_2 - 5175.
\end{aligned}
\]
Solving this system modulo $n$, one obtains $(D_0, D_1, D_2) = (7891, 7055, 9071)$, as expected.


\section{Suggested parameters and block size}\label{parameters}

Let $d_1, d_2, d_3, d_4 \in \mathbb{N}$. Let\[G(x, y) = \sum_{i=0}^{d_1} \sum_{j=0}^{d_2} a_{i,j} x^i y^j\quad \text{and}\quad H(x, y) = \sum_{i=0}^{d_3} \sum_{j=0}^{d_4} b_{i,j} x^i y^j,\] where $a_{i,j}$ are integers modulo $n$ for $i = 0,\ldots,d_1$ and $j = 0,\ldots,d_2$, and $b_{i, j}$ are integers modulo $n$ for $i = 0,\ldots,d_3$ and $j = 0,\ldots,d_4$. 

The suggested values of parameters are: $n$ is a prime of size about 20 bits, $d_1 = d_2 = d_3 = d_4 = 2$, $d = 20$. With these values of $d_1,d_2,d_3,d_4, d$, performing the proposed attacks seems impossible in practice, as we shall see in the experiments, Section \ref{Experiments}.

The values of $a_{i,j}$ can be chosen as random nonzero integers modulo $n$ for $i = 0,\ldots,d_1$ and $j = 0,\ldots,d_2$, and similarly for $b_{i, j}$ for $i = 0,\ldots,d_3$ and $j = 0,\ldots,d_4$. The values of $t_0,\ldots,t_d$ can be chosen as random nonzero integers modulo $n$.

With the suggested parameter values the size of the prefixes, contents, and suffixes, which is roughly equal to $(d+1) \log n$, is about 400 bits.

\section{Experimental results}\label{Experiments}

In this section, we show how the attacks on the basic construction may be successful but turn out to be unsuccessful when used against the advanced quantum-safe construction.

Experiments in this project were performed on Intel\textsuperscript{\textcopyright} i9-13900 processor with 50 GB of memory using 24 cores. As the main platforms, we used Maple 2024~\cite{Maple}, and two Julia packages: Groebner.jl 0.8.3~\cite{Groebner} and AlgebraicSolving.jl 0.8.2~\cite{AlgebraicSolving}. We measured memory consumption using the {\tt time} command. The code of our experiments is available on GitHub \cite{GitHub}.

\medskip


\subsection{Attack 1.}

\subsubsection{Attack on the basic construction.}

Fix a prime $n$.
Let $d \in \mathbb{N}$, and let $P_i(t), C_i(t), S_i(t)$, and $P_{i+1}(t)$ be univariate polynomials of degree $d$ with integer coefficients modulo $n$. Assume that $d < n$. Let $t_0, \ldots, t_d$ be distinct integers modulo $n$.


The integrity condition is:
\[F(P_i(t_j), C_i(t_j), S_i(t_j)) = P_{i+1}(t_j), ~j = 0,\ldots,d.\]
To change the content from $C_i(t)$ to $C'_i(t)$, one needs to find a suffix $S'_i(t)$ that satisfies
\[F(P_i(t_j), C'_i(t_j), S'_i(t_j)) = P_{i+1}(t_j), ~j = 0,\ldots,d.\]

One option is to search for a solution in the form of a polynomial with undetermined coefficients, $\hat{S'_i}(t) = \sum_{i=0}^{d} A_{i} t^i$ (the so-called ansatz polynomial), where $\bar{A} = (A_0, \ldots, A_d)$ are the new indeterminates:
\[F(P_i(t_j), C_i'(t_j), \hat{S}_i'(t_j)) = P_{i+1}(t_j), ~j = 0,\ldots,d.\]
This yields a system of $d+1$ polynomial equations in $d+1$ unknowns.

The computational hardness of solving such a system depends on the choice of secret polynomials. Let $d_1, d_2, d_3, d_4 \in \mathbb{N}$. Let the secret polynomials be \[G(x, y) = \sum_{i=0}^{d_1} \sum_{j=0}^{d_2} a_{i,j} x^i y^j\quad \text{and}\quad H(x, y) = \sum_{i=0}^{d_3} \sum_{j=0}^{d_4} b_{i,j} x^i y^j,\] where $a_{i,j}$ are integers modulo $n$ for $i = 0,\ldots,d_1$ and $j = 0,\ldots,d_2$, and $b_{i, j}$ are integers modulo $n$ for $i = 0,\ldots,d_3$ and $j = 0,\ldots,d_4$.


The degrees $d_1, d_3, d_4$ affect only the size of the coefficients that are taken modulo $n$, hence $d_1, d_3, d_4$ do not affect the computational hardness of solving the system. We fix $d_1 = d_3 = d_4 = 2$.  The coefficients of $G(x, y)$ and $H(x, y)$ can be selected from a subset of integers modulo $n$, which typically produces a zero-dimensional system of polynomial equations. 

It appears that the number of solutions of the system (in an algebraic closure of the ground field) is $d_2^{d+1}$, which coincides with the B{\'e}zout bound.
The bit complexity of Gr{\"o}bner bases for zero-dimensional systems is $(D^N)^{O(1)}$, where $D$ is the degree of polynomials and $N$ is the number of variables (see \cite{lazard} and references therein). By the B{\'e}zout theorem, this upper bound is essentially sharp. Since the number of solutions equals the bound in the B{\'e}zout theorem, we expect that the above complexity is attained for our systems.

To assess the computational hardness of solving a system of polynomial equations  experimentally, we use Gr{\"o}bner bases to solve particular instances of our system. We fix $n =11587$ and pick the coefficients of $G(x, y)$ and $H(x, y)$ as random integers in the range from $0$ to $n-1$. In Table \ref{table:attack-1-2}, we list the memory consumption of Gr{\"o}bner basis computation for different $d_2, d$. We do not report the running times since they are qualitatively similar to the memory consumption. We used the {\tt groebner} command in the Julia package Groebner.jl. We observe that incrementing either $d_2$ or $d$ by $1$ roughly doubles the memory consumption.

To test a wider selection of algorithms and implementations, we repeated the experiments using two other software packages: Maple (that, as far as we know, uses the F4 algorithm), and AlgebraicSolving.jl, which has a top-of-the-line implementation of a signature-based algorithm. These experiments led to similar conclusions.

\begin{table}
\centering
\caption{The memory consumption of a Gr{\"o}bner basis solver for the square polynomial system for basic construction in Attack 1. OOM stands for out of memory ($>50$ GB).}
\begin{tabular}{cc|rrrrrr}
    \multicolumn{1}{c}{} & \multicolumn{1}{c}{} & \multicolumn{6}{c}{$d$}\\
    & & 4 & 5 & 6 & 7 & 8 & 9\\ \cline{2-8}
    \multirow{8}{*}{\rotatebox{90}{$d_2$}}
    & 3  & 0.7 GB & 0.7 GB & 0.8 GB & 2.0 GB & 10.3 GB & OOM \\
    & 4  & 0.7 GB & 1.1 GB & 6.4 GB & OOM & - & - \\
    & 5  & 0.9 GB & 5.0 GB & OOM & - & - & - \\
    & 6  & 1.8 GB & 28.5 GB & - & - & - & - \\
    & 7  & 4.6 GB & OOM & - & - & - & - \\
    & 8  & 12.3 GB & - & - & - & - & - \\
    & 9  & 33.0 GB & - & - & - & - & - \\
    & 10 & OOM & - & - & - & - & - \\
\end{tabular}
\label{table:attack-1-2}
\end{table}

While experiments suggest that solving the above system is hard, a critical issue of the basic construction is the possibility to produce arbitrarily many more equations by specializing the integrity condition at different points.
Let $\ell \in \mathbb{N}$ with $\ell < n$, such that $d < \ell$. Let $t_0, \ldots, t_\ell$ be distinct integers modulo $n$. Consider the following (overdetermined) system:
\[
F(P_i(t_j), C'_i(t_j), \hat{S'_i}(t_j)) = P_{i+1}(t_j),  ~j = 0,\ldots,\ell.
\]
For $\ell = d + 1$, in our experiments, such system has a single solution. Taking more equations ($\ell = 2d,3d,\ldots$) significantly reduces the running time and memory consumption of the Gr{\"o}bner basis solver, which is to be expected. Thus, in practice, it is possible to solve the system rather efficiently for the basic construction. 

\subsubsection{Failure of Attack 1 on the advanced construction.}

In case of the construction from Section~\ref{sec:advanced}, the polynomial $F(x,y,z)$ is not public. Therefore, it is impossible to produce arbitrarily many polynomial equations, and one must solve a square system, which is infeasible as we have seen before. The computational hardness of solving such a system is illustrated in Table~\ref{table:attack-1-2}. From this table, we expect the memory consumption to roughly double when incrementing $d$ by $1$, and, by extrapolation, solving for $d>30$ would require at least $40$ petabytes of memory.

We note that interpolating $F(\hat{P}_i(t),\hat{C}_i(t),\hat{S}_i(t)) - \hat{P}_{i+1}(t)$ as a polynomial in $t$ of degree $d$ is impossible whenever $G(x,y), H(x,y)$ are not linear.


\subsection{Attack 2.}

\subsubsection{Attack on the basic construction.}

Fix a prime $n$.
Let $d_1, d_2, d_3, d_4 \in \mathbb{N}$. 
Let the secret polynomials be \[G(x, y) = \sum_{i=0}^{d_1} \sum_{j=0}^{d_2} a_{i,j} x^i y^j\quad \text{and}\quad H(x, y) = \sum_{i=0}^{d_3} \sum_{j=0}^{d_4} b_{i,j} x^i y^j,\] where $a_{i,j}$ are integers modulo $n$ for $i = 0,\ldots,d_1$ and $j = 0,\ldots,d_2$, and $b_{i, j}$ are integers modulo $n$ for $i = 0,\ldots,d_3$ and $j = 0,\ldots,d_4$.

Introduce new indeterminates $A_{i,j}$ for $i = 0,\ldots,d_3$ and $j = 0,\ldots,d_4$ and $B_{i, j}$ for $i = 0,\ldots,d_1$ and $j = 0,\ldots,d_2$. In case $d_1, d_2, d_3, d_4$ are known, the polynomials $G(x,y)$ and  $H(x,y)$ can be recovered by constructing the ansatz polynomials \[\hat{G}(x, y) = \sum_{i=0}^{d_1} \sum_{j=0}^{d_2} B_{i,j} x^i y^j,\quad \hat{H}(x, y) = \sum_{i=0}^{d_3} \sum_{j=0}^{d_4} A_{i,j} x^i y^j\] and solving the following system for the undetermined coefficients $A_{i,j}, B_{i, j}$:
\[
F(P_i, C_i, S_i) = \hat{G}(P_i, \hat{H}(P_i, C_i) - S_i).
\]


The dimension of such systems in all experiments we tried is 1. This can be explained by the following symmetry in the equations: if $G(x,y)$ and $H(x,y)$ satisfy the equality
$$F(x,y,z) = G(x, H(x,y)-z),$$
then $G(x,y - U)$ and $H(x,y) + U$ satisfy it as well for any $U$.
By normalizing one of the secret polynomials (say, $H(x,y)$) to have the trailing coefficient equal to one, the system becomes zero-dimensional
in all examples that we have tested.

For example, for $n = 11587$, for $d_1 = d_2 = d_3 = d_4 = 1$, for a particular choice of secret polynomials, one possible instance of the system of equations in the undetermined coefficients is:
\[
\begin{array}{rr}
   A_{0,1} B_{0,1} + 842 = 0   &  A_{1,1} B_{1,1} + 2023 = 0  \\
   11586 A_{1,1} + 5373 = 0  & A_{1,1} B_{0,0} + A_{0,1} B_{1,0} + A_{1,0} + 6818 = 0\\
   11586 A_{0,1} + 10712 = 0 &  A_{0,1} B_{0,0} + A_{0,0} + 8216 = 0\\
   A_{1,1} B_{1,0} + 289 = 0 & A_{1,1} B_{0,1} + A_{0,1} B_{1,1} + 986 = 0\\
\end{array}
\]
The system contains trivial linear equations that are readily solved, and substituting the solutions of linear equations into other equations would produce new linear equations, at which point the procedure can be repeated. This example system can be solved by such procedure.
An issue with the basic construction is that it seems that this iterative procedure can be applied for any $d_1 \geqslant 1, d_2 \geqslant 1, d_3 \geqslant 1, d_4 \geqslant 1$, which makes it straightforward to solve the resulting systems.

\subsubsection{Failure of Attack 2 on the advanced construction.}

The attack cannot be applied since an unauthorized party has no access to $F(x,y,z)$.


\subsection{Attack 3.}

\subsubsection{Attack on the basic construction.}

Direct application of~\cite[Algorithm 4]{kozen} to $F(x,y,z)$ considered as a polynomial in $\mathbb{Q}(x,z)[y]$ (that is, using $y$ as the main variable) succeeds. The output of the algorithm are $G(x,y)$ and $H(x,y)$, possibly normalized to have trailing coefficients equal to zero. As we have seen, having this output is equivalent to having a trapdoor to the construction. The complexity of the algorithm is polynomial in the degree of $F(x,y,z)$ in $y$, which makes it efficient at breaking our basic construction.

\subsubsection{Failure of Attack 3 on the advanced construction.}

In this scenario, the polynomial $F(x,y,z)$ is not available publicly. Therefore, direct application of~\cite[Algorithm 4]{kozen} is impossible.

\subsection{Attack 4.}

\subsubsection{Attack on the basic construction.}
If an attacker can observe the same block before and after it was changed, then (s)he can be in possession of the following equation:
$S_i - S_i' = H(P_i, C_i) - H(P_i, C_i') \Mod{(t-t_0) \cdots (t-t_d)}$. Here $S_i, S_i', P_i, C_i, C_i'$ are all known to the attacker.

Then, if the bound on the degree of $H(x,y)$ is known to the public (as it should be), the above  equation translates into a system of linear equations in the coefficients of $H(x,y)$. Upon accumulating sufficiently many equations as above (after sufficiently many replacements of $C_i$ by different $C_i'$), the attacker can find all the coefficients of $H(x,y)$.

\subsubsection{Failure of Attack 4 on the advanced construction.}

In the advanced construction, the attacker no longer sees the actual equality

$S_i - S_i' = H(P_i, C_i) - H(P_i, C_i') \Mod{(t-t_0) \cdots (t-t_d)},$

\noindent but knows only that there exist secret evaluation points $t_0,\ldots,t_d$ at which the corresponding equalities hold.

\section{Conclusions}

We proposed a new quantum-safe and lightweight construction for redactable  blockchains based on multivariate polynomial equations.

We studied several possible attacks on our scheme using various Gr\"obner bases computation techniques and determined that none of them is feasible against our construction.

\vspace{10pt}
\paragraph{{\bf Acknowledgements.}}
We are grateful to the referee for pointing out Attack 4.

\bibliographystyle{amsplain}

\begin{thebibliography}{1}


\bibitem{Giuseppe}
G. Ateniese, B. Magri, D. Venturi, E. Andrade,  {\it Redactable
blockchain  – or – rewriting history in bitcoin and friends}, in:  2017 IEEE European Symposium on Security and Privacy, INSPEC Accession Number: 17011479. (See also \url{https://eprint.iacr.org/2016/757.pdf})

\bibitem{Maple}
L. Bernardin, P. Chin, P. DeMarco, K. O. Geddes, D. E. G. Hare, K. M. Heal, G. Labahn, J. P. May, J. McCarron, M. B. Monagan, D. Ohashi, S. M. Vorkoetter, {\it Maple Programming Guide}, Maplesoft, a division of Waterloo Maple Inc., 1996-2023.

\bibitem{AlgebraicSolving}
J. Berthomieu, C. Eder, M. Safey El Din, {\it msolve: A Library for Solving Polynomial Systems}, Proceedings of the 46th International Symposium on Symbolic and Algebraic Computation (2021), 51--58.

\bibitem{Groebner}
A. Demin, S. Gowda, {\it Groebner.jl: A package for Gr\"obner bases computations in Julia}, preprint (2024), \url{https://arxiv.org/abs/2304.06935}.

\bibitem{Thyagarajan}
D. Deuber, B. Magri, S. A. K. Thyagarajan, {\it Redactable blockchain in the permissionless setting}, in: 2019 IEEE Symposium on Security and Privacy, 124--138. (See also \url{https://arxiv.org/abs/1901.03206})

\bibitem{NYT}
{\it  Downside  of  bitcoin:   A  ledger  that  can't  be  corrected}, The New York Times,  2016. \url{https://tinyurl.com/ydxjlf9e}

\bibitem{F4}
J.-C. Faugére, {\it A new efficient algorithm for computing Gröbner bases (F4)}, J. Pure Appl. Algebra, {\bf 139} (1999) 61--88.

\bibitem{MultivariateFunctionDecomposition}
J.-C. Faugére, {\it An efficient algorithm for decomposing multivariate polynomials and its applications to cryptography}, J. Symb. Comput. {\bf 44} 1676--1689

\bibitem{zur}
J. von zur Gathen, I. E. Shparlinski, {\it Generating safe primes}, J. Math. Cryptol. {\bf 7} (2013), 333--365.


\bibitem{GitHub}
GitHub repository,
\url{https://github.com/sumiya11/Redactable_blockchains_and_polynomial_equations}.

\bibitem{GS}
D. Grigoriev and  V. Shpilrain, {\it RSA and
redactable blockchains}, Int. J. Computer Math.: Computer Systems Theory {\bf 6} (2021), 1--6.

\bibitem{lazard}
A. Hashemi and D. Lazard, {\it Sharper Complexity Bounds for Zero-dimensional Gr{\"o}bner Bases and Polynomial System Solving}, International Journal of Algebra and Computation {\bf 21} (2011) 703--713.


\bibitem{kozen}
D. Kozen, S. Landau, {\it Polynomial decomposition algorithms}, J. Symb. Comput. {\bf 7} (1989), 445--456

\bibitem{Puddu}
I. Puddu, A. Dmitrienko, S. Capkun, {\it $\mu$chain: How to forget without hard forks}, preprint, https://eprint.iacr.org/2017/106.pdf

\bibitem{Shor}
P. Shor, {\it Polynomial-time algorithms for prime factorization and discrete logarithms on a quantum computer}, SIAM J. Comput. {\bf 26} (1997), 1484--1509.



\end{thebibliography}

\end{document}